\documentclass[6pt]{article}
\usepackage{indentfirst}
\usepackage{amsmath}
\usepackage{graphicx}
\usepackage{hyperref}
\usepackage{amssymb}

\author{Stanislav Srednyak}

\title{Universal deformation of particle momenta space in perturbation theory}
\begin{document}
\maketitle

\abstract{
We define an embedding of the space of complex momenta and masses in perturbation theory into a universal projective space. This embedding is natural in the sense of properties of the vector bundle defined by Feynman integrals on the complement to Landau varieties. We point out that there is a holonomic D-module associated with individual Feynman integrals. We quote explicit generators for this D-module on the fully deformed space of particle momenta. This basis is quadratic in the derivatives. We conclude that there is holonomic D-module on the physical space of momenta. 
}

\section{Introduction}

Particle configuration spaces are of utmost importance for both collider phenomenology and theoretical investigations in the Standard Model physics. Properties of multiparton configuration spaces are essential for investigation of confinement problem and in models of bound states. With the current formulation of quantum field theory, the only available framework for investigation is based on perturbation theory. From the pragmatic point of view of perturbation theory, it is important to analyse parton momenta configurations from the viewpoint of algebraic geometry.

In this note we make the observation that there is natural universal space in which tuples of particle momenta are embedded. These spaces are complex projective spaces of large dimension, and the space of particle momenta together with masses embed as algebraic varietiy (we make analytic continuation and consider these quantities as complex). The advantage of this viewpoint comes from the fact that the original Feynman integral can be considered as a pull-back of a flat bundle on this univesal space to the subvariety. The flat bundle on the univesal space has particularly simple form. It is a generalized hypergeometric function in the sense of Gelfand, Kapranov and Zelevinsky (GZK in the following) ~\cite{GZK_toric}. 

Apart from bringing simplicity to Feynman integrals, our appoach brings the extensive machinery developed for the study of hypergemetric functions (HG in the following). Among the many developments, we wish to point out the fact that HG functions have bases of solutions indexed by points in the secondary polytope of the integrand ~\cite{GZK_toric,Sturmfels_Grobner}. It is quite likely that, after lifting regularization, this structure will explain the emergence of polylogarithms that is typical in perturbation theory ~\cite{Arkani-Hamed}.

\section{Definition of the problem}

The classical Feynman integrals are defined as 
\begin{equation}
I(p_i,m_i)=\int_{\zeta} \prod ((q_i+p_{i,\alpha})^2+m^2_{i,\alpha})^{-1}q^\gamma d^dq_1...d^dq_L
\end{equation}
see ~\cite{Weinzierl} for details. The integration is performed over a contour with boundary at infinity in the complex projective space. We note here that we used $i\epsilon$ prescription to do complex continuation, so both momenta and masses are now in complex space. The contour of integration no longer intersects any of the Feynman quadrics
\begin{equation}
[\zeta] \in H_d(\mathbb{C}P^d-\cup_{i,\alpha}\{D_{i,\alpha}=0\},\mathbb{C}P^{d-1})
\end{equation}
This integral has therefore a vector index $\zeta$, and these intergals, collected for all values of $\zeta$, form a vector bundle. This bundle is defined for generic complex values of momenta and masses. As the parameters vary, there are topological transitions in multiple intersections of the Feynman quadrics. These values constitute Landau varieties of the integral, which form a multicomponent singularity locus of the integral ~\cite{Connes-Marcolli}. The filber bundle is flat, but has nontrivial monodromy.

\section{The universal deformation of multimomenta space}

One of the starting points of the theory of hypergeometric functions are so-called Euler integrals
\begin{equation}
I(p_{i,\omega})=\int \prod_i P_i^{\alpha_i} q^\gamma d^q.
\end{equation}

In this expression
\label{eq:Euler}
\begin{equation}
P_i=\sum_{\omega \in A_i} p_{i,\omega} q^\omega
\end{equation}
are polynomials with generic coefficients. Therefore, these integrals are naturally considered on the affine space
\begin{equation}
\{p_{i,\omega}\}=\mathbb{C}^{A_1}\otimes...\otimes \mathbb{C}^{A_s}
\end{equation}
The integrals develop singularities on the discriminants associated to the system of polynomials $P_i$. On the complement to the discriminants the integral defines a section of a flat vector bundle. The flat connection  is the Gauss-Manin connection ~\cite{GM_for_GZK}. 

It is therefore natural to define the maximal deformation by introducing variable coefficients for all the monomials involved in the denominators of the Feynman quadrics. 

{\bf Definition} Universal deformation of the multimomenta space $\{(p_{i,\alpha,\mu},m_{i,\alpha})\}$ is obtained by introducing variable coefficients for all monomials in the integrand. As a result, we obtain the following integral
\begin{equation}\label{eq:universal}
I(A_{i,\alpha,a,b,\mu},B_{i,\alpha,a,\mu},C_{i,\alpha})=\int \prod (A_{i,\alpha,a,b,\mu}q_{a,\mu}q_{b,\mu}+B_{i,\alpha,a,\mu}q_{a,\mu}+C_{i,\alpha})^{\alpha_i}q^\gamma d^dq
\end{equation}
We also introduced "fully regularized" integrals, by allowing the exponents to be arbitrary complex numbers.

It is known ~\cite{Sturmfels_Grobner} that the HG functions may have poles at integer powers $\alpha_i, \gamma_\mu$. This fact corresponds to degeneration in the cohomology of the local system defined by the integrand. This degeneration is very important as it is related to the mechanism by which polylogarithms emerge. However, the theory is much simpler for generic complex values of the powers $\alpha_i$ and in this paper we restrict our attention only to this case. 

We now turn to the description of the embedding of physical space of momenta into the universal space. It is given by the choice
\begin{gather*}
A_{i,\alpha,a,b,\mu}=0\ or\ 1 \\
B_{i,\alpha,a,\mu}=2p_{i,\alpha,a,\mu}\\
C_{i,\alpha}=p_{i,\alpha}^2+m^2_{i,\alpha}
\end{gather*}

\section{Holomorphic properties of the vector bundle on the universal space}

The integrals ~\ref{eq:universal} define a vector bundle on the parameter space, as for different choice of contour of integration we obtain different function. The variation of the integral as the parameters vary along a loop in the complement to the multiple dicriminantal loci
\begin{equation}
D_{_1,...,i_s}=Discr(D_{i_1},...,D_{i_s})
\end{equation}
is given by the variation matrix that acts on the middle dimensional cohomology
\begin{gather*}
Var_{i_1,...,i_s}:H^d(\mathbb{C}P^n-\{P_{i_1}=0\}\cup...\cup\{P_{i_s}=0\}) \rightarrow \\
H^d(\mathbb{C}P^n-\{P_{i_1}=0\}\cup...\cup\{P_{i_s}=0\})
\end{gather*}
It is known ~\cite{AGV_2} that this operator is nilpotent and can be related to the mixed Hodge structure on the space $\mathbb{C}P^n-\{P_{i_1}=0\}\cup...\cup\{P_{i_s}=0\}$. Algorithms for its computation were discussed in ~\cite{Khovanskii-Danilov} from the point of view of combinatorial geometry. 

The remarkable discovery of ~\cite{GZK_toric} was that there is a holonomic D-module associated to the function ~\ref{eq:Euler}. We will describe this D-module for the case of fully deformed and fully regularized Feynman integrals ~\ref{eq:universal}. The relations in the GZK D-module are of 3 types.

Type 1. Lattice relations. 

Consider a vector in the space $S=\mathbb{Z}^{A_1}\otimes...\otimes \mathbb{Z}^{A_s}$ with components $(a_{i,\omega_i})$, such that the following holds
\begin{gather*}
\sum_{i,\omega_i} a_{i,\omega_i}=0 \\
\sum_{i,\omega_i} a_{i,\omega_i}\omega_{i,\mu}=0 ,\mu=1,...d
\end{gather*}
These equations define a lattice $L$ in the space $S$. For each vector in this lattice $L$, split it into two parts $a^+_{i,\omega_i}$ and $a^-_{i,\omega_i}$ that contain positive and negative coordinates respectively. For such a splitting, consider the operator
\begin{equation}
a
\Box(a)=\prod_{i,\omega_i}(\frac{\partial}{\partial a_{i,\omega_i}})^{a^+_{i,\omega_i}}-\prod_{i,\omega_i}(\frac{\partial}{\partial a_{i,\omega_i}})^{-a^-_{i,\omega_i}}
\end{equation}
It is easy to see that each of the Euler intergals satisfies 
\begin{equation}
\Box_aI=0
\end{equation}

In practice, it is enough to have finitely many of such equations, one for each of the basis vector of the lattice $L$.

Type 2. Integration by parts (IBP) relations.

These equations come from the basic IBP identity
\begin{equation}
0=\int \frac{\partial}{\partial q_\mu} (\prod P_i^{\alpha_i} q^{\gamma+\delta_\mu})d^d q
\end{equation}
Which gives 
\begin{equation}
\sum_i \sum_{\omega_i} \omega_{i,\mu} p_{i,\omega_i}\frac{\partial}{\partial p_{i,\omega_i}}I+(\gamma_\mu+1)I=0
\end{equation}

Starting from particular (complex $\gamma_\mu$ ) we can obtain a lattice of relations.

Type 3. Homogeneity relations. 

These relations are of the form
\begin{equation}
 \sum_{\omega_i}  p_{i,\omega_i}\frac{\partial}{\partial p_{i,\omega_i}}I-\alpha_i I=0
\end{equation}

It can be proven ~\cite{GZK_toric} that these relations define a holonomic D-module. 

We observe that in the case of fully deformed Feynman integrals it is possible to decribe the lattice $L$ fully. It is generated by the following terms
\begin{gather*}
\frac{\partial}{\partial A_{i,\alpha,a,b,\mu}}  \frac{\partial}{\partial C_{j,\beta}}-\frac{\partial}{\partial B_{i,\alpha,a,\mu}}   \frac{\partial}{\partial B_{j,\beta,b,\mu}},\\
\frac{\partial}{\partial A_{i,\alpha,a,b,\mu}}  \frac{\partial}{\partial C_{j,\beta}}-\frac{\partial}{\partial B_{i,\alpha,b,\mu}}   \frac{\partial}{\partial B_{j,\beta,a,\mu}},\\
\frac{\partial}{\partial A_{i,\alpha,a,b,\mu}}  \frac{\partial}{\partial C_{j,\beta}}-\frac{\partial}{\partial B_{i,\beta,a,\mu}}   \frac{\partial}{\partial B_{j,\alpha,b,\mu}},\\
\frac{\partial}{\partial A_{i,\alpha,a,b,\mu}}  \frac{\partial}{\partial C_{j,\beta}}-\frac{\partial}{\partial B_{i,\beta,b,\mu}}   \frac{\partial}{\partial B_{j,\alpha,a,\mu}}.\\
\end{gather*}

The proof is most easily conducted by the Fourier method, pioneered in ~\cite{GZK_toric}. Here we just give a heuristic dimension counting argument. In Fourier space, the lattice relations can be solved for Fourier conjugates of $A_{i,\alpha,a,b,\mu}$, as there is at least one such equation for each of these variables. There are as many homogeneity equations as there are $C_{j,\beta}$ variables. There is certain subtlety involved in defining Fourier conjugates of $B_{i,\alpha,a,\mu}$ which is difficult to make explicit without use of toric geometry. We refer to the original paper ~\cite{GZK_toric} for accurate counting argument.

\section{Examples.}

\subsection{1-loop}

In this section we consider the simplest of the cases, namely 1-loop integrals. The starting point is
\begin{equation}
I=\int \prod ((q+p_i)^2+m_i^2)^{-1}q^\gamma d^dq
\end{equation}
The fully deformed integral takes the form
\begin{equation}
I=\int \prod (A_{i,\mu}q_\mu^2+B_{i,\mu}q_\mu +C_i)^\alpha q^\gamma d^d q
\end{equation}
Therefore, the original space $\{p_{i,\mu},m_i\}= \mathbb{C}^{s(d+1)}$ is embedded in $\mathbb{C}^{s(2d+1)}$. The lattice relations take the form
\begin{equation}
(\frac{\partial}{\partial A_{i,\mu}} \frac{\partial}{\partial C_{j}} -\frac{\partial}{\partial B_{i,\mu}}\frac{\partial}{\partial B_{j,\mu}} ) I=0
\end{equation}

\subsection{2-loop}

The starting point in this case is
\begin{equation}
I=\int \prod ((q_1+p_{1,\alpha})^2+m^2_{1,\alpha})^{-1}  ((q_2+p_{2,\alpha})^2+m^2_{2,\alpha})^{-1}  ((q_1+q_2+p_{3,\alpha})^2+m^2_{3,\alpha})^{-1} q^\gamma dq
\end{equation}

The fully deformed integral reads
\begin{gather*}
I=\int \prod (A_{1,\alpha,1,1,\mu}q_{1,\mu}^2+B_{1,\alpha,1,\mu}q_{1,\mu}+C_{1})^{\alpha_{1,\alpha}} \\
(A_{2,\alpha,2,2,\mu}q_{2,\mu}^2+B_{2,\alpha,2,\mu}q_{2,\mu}+C_{2,\alpha})^{\alpha_{2,\alpha}} \\
(A_{3,\alpha,1,1,\mu}q_{1,\mu}^2+A_{3,\alpha,1,2,\mu}q_{1,\mu}q_{2,\mu}+A_{3,\alpha,2,2,\mu}q_{2,\mu}^2+\\
B_{3,\alpha,1,\mu}q_{1,\mu}+B_{3,\alpha,2,\mu}q_{2,\mu}+C_{2,\alpha})^{\alpha_{3,\alpha}}q^\gamma dq
\end{gather*}
For each $\alpha,\beta$ there are 30 lattice equations. Below we list some of them
\begin{gather*}
(\frac{\partial}{\partial A_{1,\alpha,1,1,\mu}} \frac{\partial }{\partial C_{1,\beta}} - 
\frac{\partial}{\partial B_{1,\alpha,1,\mu}} \frac{\partial }{\partial B_{1,\beta,1,\mu}})I=0 \\
(\frac{\partial}{\partial A_{3,\alpha,1,2,\mu}} \frac{\partial }{\partial C_{1,\beta}} - 
\frac{\partial}{\partial B_{3,\alpha,1,\mu}} \frac{\partial }{\partial B_{1,\beta,2,\mu}})I=0 \\
\end{gather*}

\section{Conclusion}

In this paper we defined an embedding of the physical momenta space of a Feynman diagram in a complex projective space. This embedding is natural from the point of view of the flat vector bundle that defines the perturbative amplitude. The holomorphic system of differential equations takes especially simple form on this space. There is substantial amount of extra structure attached to this space. The hypergeometric function that defines fully regularized integral is essentially determined by its monodromy around components of the principal discriminant ~\cite{GZK_book,Borel_D_mod}. The components of the principal discriminant are essentially determined by the toric geometry of the algebraic torus defined by the integration variables
\begin{equation}
(q_{a,\mu})\in (\mathbb{C}^*)^d->(1,q_{a,\mu},q_{a,\mu}q_{b,\nu}) \in \mathbb{C}^\Omega
\end{equation}
The combinatorial information coming from representations of this torus must play important role in the structure of Feynman amplitudes. In particular, the volume of the polytope of the toric variety cited above determines the number of master integrals ~\cite{GZK_book} necessary for determining the Gauss-Manin connection, a topic of large interest in computational approaches. ~\cite{Mastrolia}

More broadly, the universal space we introduced will probably play a role in nonperturbative formulations of QCD. The reason we can express this belief is the similarity of amoebas of discrimiantal loci ~\cite{GZK_book} ( which are known to be related to the convergence regions of $\Gamma$-series expansions and therefore to the asymptotics near the discriminantal loci) and the observed events at hadronic colliders. This topic is under active investigation.

\bibliographystyle{amsplain}
\bibliography{bibliogr}

\providecommand{\bysame}{\leavevmode\hbox to3em{\hrulefill}\thinspace}
\providecommand{\MR}{\relax\ifhmode\unskip\space\fi MR }
\providecommand{\MRhref}[2]{%
  \href{http://www.ams.org/mathscinet-getitem?mr=#1}{#2}
}
\providecommand{\href}[2]{#2}
\begin{thebibliography}{10}

\bibitem{Mastrolia}
Mario Argeri, Stefano Di~Vita, Pierpaolo Mastrolia, Edoardo Mirabella, Johannes
  Schlenk, Ulrich Schubert, and Lorenzo Tancredi, \emph{{Magnus and Dyson
  Series for Master Integrals}}, JHEP \textbf{03} (2014), 082.

\bibitem{Arkani-Hamed}
Nima Arkani-Hamed, Jacob~L. Bourjaily, Freddy Cachazo, Alexander~B. Goncharov,
  Alexander Postnikov, and Jaroslav Trnka, \emph{{Grassmannian Geometry of
  Scattering Amplitudes}}, Cambridge University Press, 2016.

\bibitem{AGV_2}
V.I. Arnold, A.N. Varchenko, and S.M. Gusein-Zade, \emph{Singularities of
  differentiable maps, volume ii monodromy and asymptotic integrals},
  Birkhauser, 1988.

\bibitem{Weinzierl}
Christian Bogner and Stefan Weinzierl, \emph{{Feynman graph polynomials}}, Int.
  J. Mod. Phys. \textbf{A25} (2010), 2585--2618.

\bibitem{Connes-Marcolli}
Alain Connes and Matilde Marcolli, \emph{Noncommutative geometry, quantum
  fields and motives}, vol.~55, Colloquium Publications, 2007.

\bibitem{Khovanskii-Danilov}
V.~I. Danilov and A.~G. Khovanskii, \emph{Newton polyhedra and an algorithm for
  computing hodge–deligne numbers}, Izv. Akad. Nauk SSSR Ser. Mat. (1986).

\bibitem{Borel_D_mod}
A.~Borel et~al, \emph{Algebraic d-modules}, Academic Press, 1987.

\bibitem{GZK_toric}
I.~Gelfand, M.Kapranov, and A.~Zelevinsky, \emph{Hypergeometric functions and
  toric varieties}, Funk. anal. and Appl \textbf{23} (1989), no.~2.

\bibitem{GZK_book}
\bysame, \emph{Discriminants, resultants, and multidimensional determinants},
  Birkhauser, 1994.

\bibitem{GM_for_GZK}
Takayuki Hibi, Kenta Nishiyama, and Nobuki Takayama, \emph{Pfaffian systems of
  a-hypergeometric equations i: Bases of twisted cohomology groups},
  arxiv.org/1212.6103 (2014).

\bibitem{Sturmfels_Grobner}
Mutsumi Saito, Bernd Sturmfels, and Nobuki Takayama, \emph{Grobner deformations
  of hypergeometric differential equations}, Springer, 2000.

\end{thebibliography}

\end{document}